\documentclass[aps,prl,twocolumn,floatfix]{revtex4}
\usepackage{times}
\usepackage{graphicx}
\usepackage{epsfig}
\usepackage{amsmath}
\usepackage{amssymb}
\usepackage[version=3]{mhchem}

\begin{document}

\title{Incommensurate chirality density wave transition in a hybrid molecular framework}

\author{Joshua A. Hill}
\affiliation{Department of Chemistry, University of Oxford, Inorganic Chemistry Laboratory, South Parks Road, Oxford OX1 3QR, U.K.}

\author{Kirsten E. Christensen}
\affiliation{Department of Chemistry, University of Oxford, Inorganic Chemistry Laboratory, South Parks Road, Oxford OX1 3QR, U.K.}

\author{Andrew L. Goodwin}
\affiliation{Department of Chemistry, University of Oxford, Inorganic Chemistry Laboratory, South Parks Road, Oxford OX1 3QR, U.K.}

\date{\today}
\begin{abstract}
Using single-crystal X-ray diffraction we characterise the 235\,K incommensurate phase transition in the hybrid molecular framework tetraethylammonium silver(I) dicyanoargentate, [NEt$_4$]Ag$_3$(CN)$_4$. We demonstrate the transition to involve spontaneous resolution of chiral [NEt$_4$]$^+$ conformations, giving rise to a state in which molecular chirality is incommensurately modulated throughout the crystal lattice. We refer to this state as an incommensurate chirality density wave (XDW) phase, which represents a fundamentally new type of chiral symmetry breaking in the solid state. Drawing on parallels to the incommensurate ferroelectric transition of NaNO$_2$ we suggest the XDW state arises through coupling between acoustic (shear) and molecular rotoinversion modes. Such coupling is symmetry-forbidden at the Brillouin zone centre but symmetry-allowed for small but finite modulation vectors $\mathbf q=[0,0,q_z]^\ast$. The importance of long-wavelength chirality modulations in the physics of this hybrid framework may have implications for the generation of mesoscale chiral textures, as required for advanced photonic materials.
 \end{abstract}

\maketitle

Symmetry breaking processes that involve ordering of chiral degrees of freedom are implicated in the physics of a variety of exotic states of matter. Both the optical gyrotropy \cite{Orenstein_2013} and polar Kerr effect \cite{Xia_2006,Konig_2017} of underdoped cuprates arise from chiral ordering of loop-current anapoles \cite{Pershoguba_2013}. Likewise, chirality density wave (XDW \cite{note}) formation is implicated in the `hidden order' state of URu$_2$Si$_2$ \cite{Kung_2015}. The possibility \cite{Kallin_2009,Kallin_2012} that chiral $p$-wave symmetry characterises the superconducting state of Sr$_2$RuO$_4$ has generated intense interest because the system should then support not only Majorana fermions \cite{Nobukane_2011} but also the non-Abelian winding statistics exploitable in quantum computing \cite{Nayak_2008}. Emergent chiral phases are observed in frustrated magnets \cite{Grohol_2005,Batista_2016} and in multi-$\mathbf q$ charge density wave (CDW) states, \emph{e.g.}\ in TeSe$_2$ \cite{Ishioka_2010,vanWezel_2010,Castellan_2013}. While in all these examples chirality emerges from peculiarities of the underlying electronic structure, there is intense fundamental interest in the phenomenology of chiral order \cite{Gopalan_2011}, irrespective of its particular physical origin.

In a chemical context, chirality arises from the spatial arrangement of molecules \cite{Pasteur_1848}, supramolecular assemblies \cite{Liu_2015,Roche_2016}, or extended lattices \cite{Kepert_2000,Wells_1977}. It is characterised formally by the absence of rotoinversion symmetry, which includes inversion and mirror symmetry. In the solid state, molecular (\emph{e.g.}\ proteins) and lattice (\emph{e.g.}\ $\beta$-quartz) chirality are usually fixed since interconversion between states of opposite handedness usually requires breaking of chemical bonds. By contrast, supramolecular assemblies---held together by weak non-covalent interactions---can in favourable cases reorganise dynamically to allow reversible chiral symmetry breaking; this is the mechanism responsible for the emergence of chiral cholesteric phases in liquid crystals, for example \cite{Dierking_2014}.

A conceptually related phenomenon is that of dynamic interconversion between locally-chiral states in the metal--organic framework (MOF)  [Zn$_2$(bdc)$_2\cdot$dabco] (bdc$^{2-}$ = 1,4-benzenedicarboxylate anion; dabco = diazabicyclooctane) \cite{Gabuda_2008,Gabuda_2011}. Here chirality is associated with two equivalent conformations (left-handed twist; right-handed twist) of the dabco molecule. On cooling below 25\,K, the system appears from $^1$H NMR measurements \cite{Gabuda_2011} to undergo a transition from dynamic enantiomer interconversion (`parachiral' state) to a state in which molecule chiralities are fixed on NMR timescales, reflecting at least a local---if not necessarily long-range---chiral symmetry breaking process. A key success of Refs.~\citenum{Gabuda_2011,Gabuda_2014} was to show that framework structures incorporating molecules whose chirality arises from conformational states (\emph{i.e.}\ rather than chemical connectivity) may in principle exhibit thermally-accessible chiral ordering transitions.

\begin{figure}[b]
\begin{center}
\includegraphics{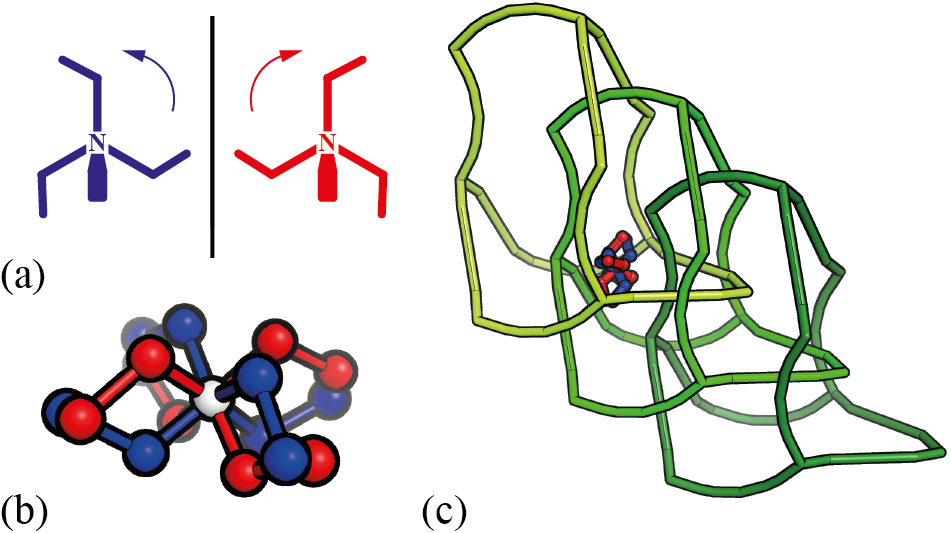}
\end{center}
\caption{\label{fig1} (a) Representations of the [NEt$_4$]$^+$ molecular geometry distinguishing left-handed (blue) and right-handed (red) conformations, related by mirror symmetry (vertical line). (b) Structural model for the [NEt$_4$]$^+$ site in the ambient-temperature phase of [NEt$_4$]Ag$_3$(CN)$_4$ \cite{Hill_2016}: the molecular cation is disordered over the two equally-populated conformations, represented here in blue and red. The two conformations are related by a mirror plane bisecting the N atom (grey). H atoms omitted for clarity. (c) The disordered molecular cations are located within cavities formed by the interpenetrating diamondoid [Ag$_3$(CN)$_4$]$^-$ frameworks (shades of green).}
\end{figure}

It was in this spirit that we chose to study the phase behaviour of the hybrid molecular framework tetraethylammonium silver(I) dicyanoargentate, [NEt$_4$]Ag$_3$(CN)$_4$ (Et = ethyl) \cite{Hill_2016}. Though achiral in both solution and the gas phase, the [NEt$_4$]$^+$ cation can nevertheless adopt a geometry in condensed phases with non-superimposable left- and right-handed conformations [Fig.~\ref{fig1}(a)]. In the room-temperature crystal structure of [NEt$_4$]Ag$_3$(CN)$_4$, both conformations co-exist in a disordered fashion: the electron density at the molecular cation site reflects an equal population of left- and right-handed conformations, with the two related by a crystallographic mirror plane [Fig.~\ref{fig1}(b)]. We reported in Ref.~\citenum{Hill_2016} the existence of an incommensurate phase transition at $T_{\rm c}=235$\,K in this system. In this Letter, we characterise this transition and show it to involve spontaneous resolution of [NEt$_4$]$^+$ conformations, giving rise to a state in which molecular chirality is incommensurately modulated throughout the crystal lattice: we describe this as an incommensurate XDW phase (by analogy to \cite{Kung_2015}) and suggest that it represents a fundamentally new type of chiral symmetry breaking in the solid state. 

Following the approach of Ref.~\citenum{Hill_2016} we prepared single crystals of [NEt$_4$]Ag$_3$(CN)$_4$. Two separate sets of single-crystal X-ray diffraction measurements were performed: the first involved data collection at 100--300\,K using an Oxford Diffraction Supernova diffractometer fitted with an Oxford Cryosystems 700 Plus open-flow nitrogen cooling device; the second involved a single data collection at 30\,K on the I19 Beamline at the Diamond Light Source, making use of an Oxford Cryosystems HeliX open-flow helium cryostat \cite{Goeta_1999}. In both cases, crystals were mounted in perfluoropolyether oil on a MiTeGen loop. The CrysAlisPro and CrystalClear software were used for data collection and reduction as appropriate \cite{Crysalis_2010,CrystalClear_2009}; structure refinement was carried out in JANA2006 \cite{Petricek_2014}.

At ambient temperature, the crystal structure of [NEt$_4$]Ag$_3$(CN)$_4$ has orthorhombic $Pnma$ symmetry \cite{Hill_2016}: three crystallographically-related diamondoid [Ag$_3$(CN)$_4$]$^-$ lattices interpenetrate, with the conformationally-disordered [NEt$_4$]$^+$ counterions situated in extra-framework cavities [Fig.~\ref{fig1}(c)]. Argentophilic interactions \cite{Jansen_1987,Schmidbaur_2015} connect adjacent frameworks, presumably stabilising the particular mode of interpenetration adopted. There is no appreciable void space in this structure; hence one anticipates strong coupling between cation orientations and framework displacements, as in \emph{e.g.}\ the organic lead halide perovskites \cite{Frost_2016}. The single crystallographically-distinct [NEt$_4$]$^+$ cation is located on the $4c$ Wyckoff position, the site symmetry of which is $.m.$\ (\emph{i.e.}, with the mirror plane lying perpendicular to the $\mathbf b$ crystal axis). The important point is that this site symmetry is incompatible with the molecular point symmetry of any chiral species.

\begin{figure}
\begin{center}
\includegraphics{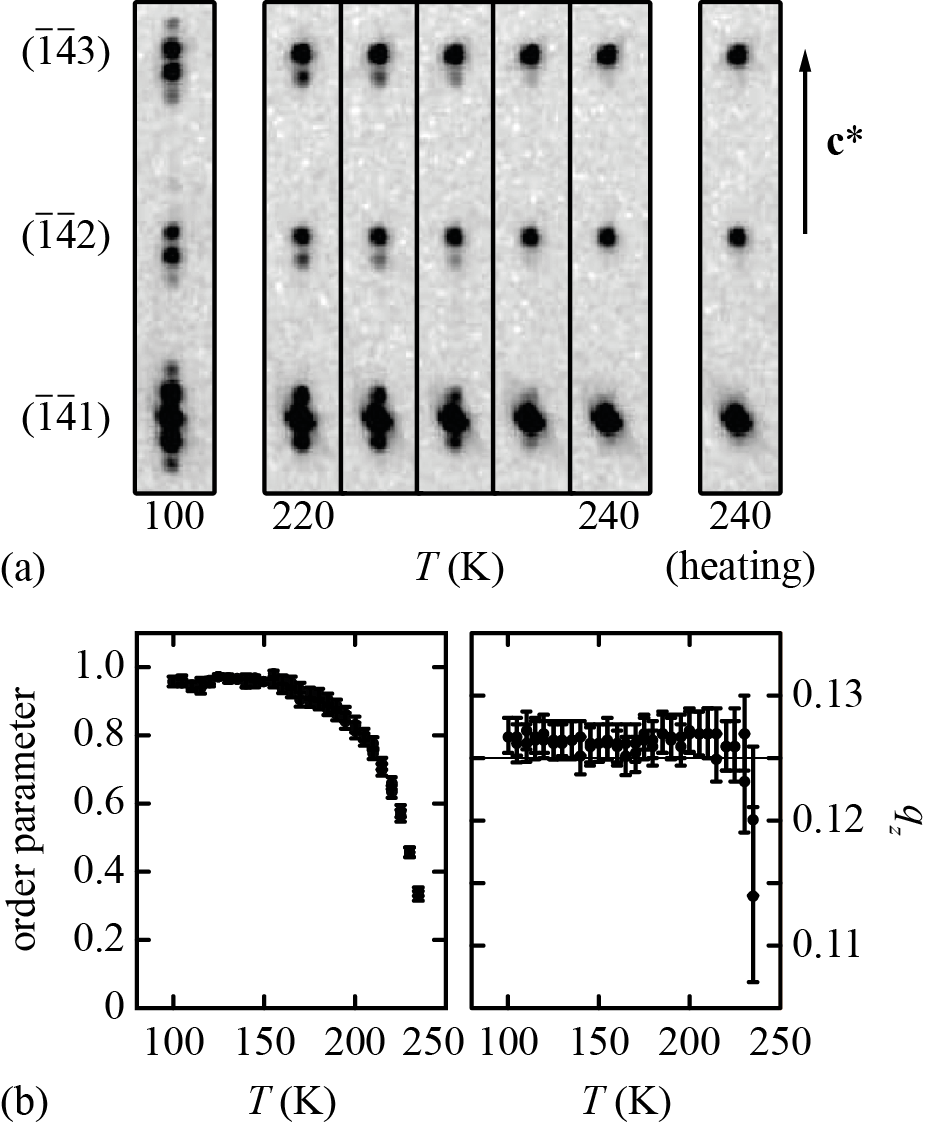}
\end{center}
\caption{\label{fig2} (a) Sections of single-crystal X-ray diffraction patterns of [NEt$_4$]Ag$_3$(CN)$_4$ collected at temperatures spanning the incommensurate phase transition. The superlattice reflections associated with incommensurate modulation of the structure are visible as satellites of the parent Bragg reflections. Note the presence at the lowest temperatures of higher-order modulation peaks ($m>1$). The transition is entirely reversible, as is evident from the similarity in diffraction patterns at 240\,K prior to and after cooling through $T_{\textrm c}$. (b) The left-hand panel shows the temperature dependence of the XDW order parameter, as determined from the relative intensity of first-order modulation reflections (averaged over six representative reflections, see SI). The right-hand panel shows the value of the modulation wave-vector component $q_z$ as a function of temperature. The nearby commensurate value $q_z=\frac{1}{8}$ is shown as a horizontal line.}
\end{figure}

On cooling below $T_{\rm c}$ we observe the emergence of a set of additional reflections associated with wave-vectors $\mathbf Q = \mathbf G+m\mathbf q$, where $\mathbf G$ is a reciprocal lattice vector, $\mathbf q$ a modulation vector $\mathbf q=[0,0,q_z]$, and $m\in\mathbb Z$ the superspace index [Fig.~\ref{fig2}(a)]. The first-order ($|m|=1$) superlattice reflection intensities vary as the square of the order parameter, the temperature dependence of which is shown in Fig.~\ref{fig2}(a). The transition is probably first-order in nature, given that there is a small discontinuity in the $c$ unit cell parameter at $T_{\rm c}$ (see SI). The modulation wave-vector component $q_z$ is evidently independent of temperature [Fig.~\ref{fig2}(b)]: we find $q_z=0.12867(16)$ for our 100--300\,K data set and $q_z=0.1278(3)$ for the 30\,K data set (the variation in $q_z$ here is within the uncertainty expected for data collected using different diffractometers and different crystals \cite{Cummins_1990}). Temperature-independent modulation wave-vectors are not uncommon amongst incommensurate phase transitions: notable examples are the transitions in quartz, NaNO$_2$, and biphenyl \cite{McConnell_1991,Cummins_1990}. So our experimental measurements indicate a low-temperature modulation of the [NEt$_4$]Ag$_3$(CN)$_4$ crystal structure by a set of distortions that repeat along the $\mathbf c$ crystal axis with wavelength $c/q_z\simeq7$\,nm.

The particular superlattice reflection conditions of the low-temperature diffraction pattern of [NEt$_4$]Ag$_3$(CN)$_4$ unambiguously identify the low-temperature superspace group as $Pnma(00\gamma)0s0$. Crucially, the symmetry-lowering process described by a transition from $Pnma$ to this superspace group removes precisely those mirror symmetry elements incompatible with chiral order at the [NEt$_4$]$^+$ site and replaces them with superspace ($s$-)glides in the same orientation. We will come to show that key consequence is that the two enantiomeric [NEt$_4$]$^+$ conformations are no longer disordered throughout the crystal, but are instead modulated by the incommensurate wave-vector $\mathbf q$: the $s$-glides map right- and left-handed conformations onto one another.

\begin{figure}
\begin{center}
\includegraphics{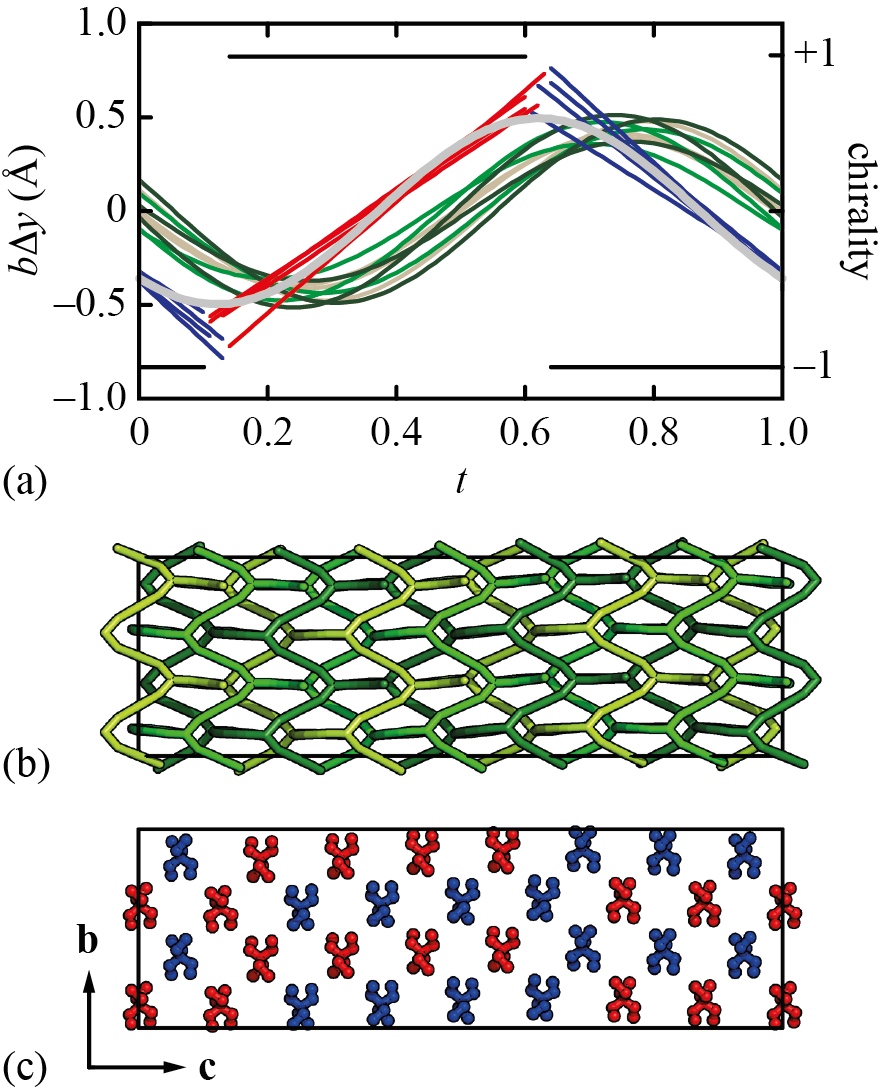}
\end{center}
\caption{\label{fig3} Incommensurate modulation in [NEt$_4$]Ag$_3$(CN)$_4$ determined using single-crystal X-ray diffraction measurements at 100\,K. (a) Displacement amplitude (coloured lines) and [NEt$_4$]$^+$ chirality as a function of modulation coordinate $t$. Displacements of the framework components are shown in shades of green; those of the central N atom in the [NEt$_4$]$^+$ cation are shown in grey; and those of the pendant C atoms of the [NEt$_4$]$^+$ cations are shown in blue or red, depending on the particular chiral state present for corresponding $t$ value. (b) Representation of the framework modulation in the commensurate approximant of the XDW state. Note that the framework displacements are predominantly polarised in the $\mathbf b$ direction. (c) Representation of the modulation in [NEt$_4$]$^+$ cation positions and chirality states (blue or red) in the same approximant model.}
\end{figure}

Superstructure refinement using the data collected at 100\,K reveals two dominant contributions to the structural modulation activated at $T_{\rm c}$: (i) a continuous displacement modulation of both framework and [NEt$_4$]$^+$ counterions that is polarised predominantly along the $\mathbf b$ crystal axis, and (ii) a crenel-type occupancy modulation of the two enantiomorphic conformations of the [NEt$_4$]$^+$ ion. The amplitudes of these two components as a function of modulation coordinate are shown in Fig.~\ref{fig3}(a). We note that the displacement magnitude is relatively large---a shift of \emph{ca} 1\,\AA\ between maximum and minimum $b\Delta y$ values---and is best modelled using sinusoidal and saw-tooth modulation functions for the framework and cation, respectively (see SI for further discussion and full details of our refinements). But our key result is the incommensurate modulation of chirality throughout the [NEt$_4$]Ag$_3$(CN)$_4$ structure; this is the XDW state, which to the best of our knowledge has not previously been reported. Because $q_z\simeq\frac{1}{8}$ we can generate an approximant of the modulated structure with $\mathbf c_{\textrm{XDW}}\simeq8\mathbf c_{Pnma}$. Graphical representations of the 100\,K framework displacements, [NEt$_4$]$^+$ molecular ion displacements, and [NEt$_4$]$^+$ enantiomeric states present in this approximant model are shown in Fig.~\ref{fig3}(b,c); the model itself is included as Supporting Information.

A transition with close conceptual similarity occurs in the inorganic salt sodium nitrite, NaNO$_2$ \cite{Yamada_1963,Heine_1984}. The NO$_2^-$ ion has a bent geometry and so supports a local dipole moment. At temperatures above $T_{\textrm{IC}}=436$\,K, NaNO$_2$ is in a paraelectric state, with NO$_2^-$ ion orientations disordered over two positions corresponding to inverse polarisation states. The system begins to order on cooling below $T_{\textrm{IC}}$, but enters an incommensurate regime before settling into a ferroelectric state at $T_\textrm{C}=434$\,K. The crystal symmetry of NaNO$_2$ is such that chiral degrees of freedom do not play a role in its phase transition behaviour. Nevertheless the two polarisation states of the NO$_2^-$ ion act as an Ising variable (or pseudospin) in precisely the same way as do the two chirality states of [NEt$_4$]$^+$; both local degrees of freedom are odd under inversion, and hence the elegant body of symmetry analysis developed to understand NaNO$_2$ \cite{Heine_1981,LyndenBell_1994} is directly applicable to [NEt$_4$]Ag$_3$(CN)$_4$.

The key rationale for an incommensurate phase transition in systems such as NaNO$_2$ is the interplay of two structural distortion modes $\psi,\phi$ of different symmetries. This is the `structrual resonance' mechanism of Heine and McConnell \cite{Heine_1984b}, which is applicable not only to NaNO$_2$ but also to a diverse range of systems including minerals \cite{Levanyuk_1976}, molecular solids \cite{Bak_1976}, and alloys \cite{Friedel_1977}. The central concept is that modulation by some wave-vector $\mathbf q$ allows the two distortion modes to interact; moreover for degrees of freedom with odd parity (such as dipoles or chiral states) the strength of this interaction scales as $|\mathbf q|^2$ and so becomes large for some small but finite $\mathbf q\neq[0,0,0]$. A characteristic signature of the mechanism is a phase shift of \emph{ca} $\pi/2$ in the modulation of the two structural distortions, which is a consequence of the role played by the Lifshitz invariant
\begin{equation}
(\nabla\psi)\phi-\psi(\nabla\phi)
\end{equation}
in the free energy expansion \cite{McConnell_1991,Lifshitz_1942}. In the case of NaNO$_2$ the two distortions are (i) shear and (ii) ferroelectric order of NO$_2^-$ ions.

We suggest that the incommensurate XDW state in [NEt$_4$]Ag$_3$(CN)$_4$ can be rationalised using entirely analogous arguments. In the $\mathbf k\rightarrow\mathbf 0$ limit, the framework modulation shown in Fig.~\ref{fig3}(b) maps onto a $yz$ shear distortion. This distortion mode is characterised by the $\Gamma_3^+$ irreducible representation (irrep.)\ and, if active, would reduce the space-group symmetry of the parent structure from $Pnma$ to $P\frac{2_1}{n}11$ ($\equiv P2_1/c$) \cite{Stokes_1988}. Group-theoretical analysis using the ISOTROPY software \cite{Campbell_2006} shows the $\mathbf k\rightarrow\mathbf 0$ limit of the chiral ordering pattern illustrated in Fig.~\ref{fig3}(c) to be characterised by the $\Gamma_4^-$ irrep. If activated in isolation, this second distortion would reduce the parent space-group symmetry to $Pn2_1a$ ($\equiv Pna2_1$). Hence the two distortions are symmetry-forbidden from mixing in a commensurate state. For modulation wave-vectors of the type $\mathbf q=[0,0,\xi]^\ast$, both distortions transform as the $\Lambda_3$ irrep., and so can couple in accordance with the theory outlined in Refs.~\citenum{Heine_1984,Heine_1984b}. The $\Lambda_3$ irrep.\ can be shown to act as the primary order parameter for  descent of symmetry from $Pnma$ to $Pnma(00\gamma)0s0$ \cite{Stokes_2013}. Furthermore, the empirical phase shift between the modulation of [NEt$_4$]$^+$ chirality and framework displacement is close to $\pi/2$ [\emph{i.e.}, $\Delta t\simeq\frac{1}{4}$ in Fig.~\ref{fig3}(a)].

So the incipient chiral order in ambient-phase [NEt$_4$]Ag$_3$(CN)$_4$ is relatively straightforward and involves a tendency to form alternate chains of right- and left-handed [NEt$_4$]$^+$ conformations. But the interplay of this ordering motif with a shear instability leads to an incommensurate XDW state, essentially as a consequence of crystal symmetry.

One might then anticipate that [NEt$_4$]Ag$_3$(CN)$_4$ may not be unique amongst molecular frameworks in exhibiting such a state: shear instabilities are relatively common for this broader family, primarily as a result of the low energy penalty associated with framework deformation (see \emph{e.g.}\ \cite{Cairns_2013}). In designing XDW systems one obvious requirement is the presence of an optically-active molecular component, which imposes its own size constraints. Hence XDW states are more likely to be found in open framework structures, such as MOFs \cite{Hoskins_1990,Rowsell_2004} or their inorganic counterparts \cite{Hill_2016,Li_2017}, capable of containing chiral molecules within their framework cavities. In such cases, the onset temperature of the XDW state will be a function of the barrier height for enantiomeric interconversion---and hence molecules with chiral states that arise from conformation rather than topology will favour XDW transitions at accessible temperatures since interconversion does not require bond breaking.

From a conceptual viewpoint the physics at play in [NEt$_4$]Ag$_3$(CN)$_4$ shares much in common with the established phenomenology of translation--rotation coupling in molecular solids \cite{LyndenBell_1994}. The key difference is the thermal accessibility of internal degrees of freedom beyond molecular translations and rotations alone. While a variety of different types of conformational degrees of freedom are very possibly relevant in many hybrid frameworks (\emph{e.g.}\ staggered--eclipsed NH$_3$CH$_3^+$ conformations in methylammonium lead iodide \cite{Whitfield_2016}), here we have focused on chirality because rotoinversion is the obvious point symmetry operation absent from the `classical' systems of Ref.~\citenum{LyndenBell_1994}. The ability to modulate chirality over nanometre lengthscales suggests an attractive bottom-up approach for developing next-generation photonic materials that exploit the interaction between chiral matter and circularly polarised light \cite{Lodahl_2017}. Moreover, there is genuine scope for control: taking [NEt$_4$]Ag$_3$(CN)$_4$ as an example, the modulation wavelength $\lambda=c/q_z$ will be affected by the shear modulus of the $[$M$_3$(CN)$_4]^-$ framework, which in turn could be tuned by chemical substitution \emph{e.g.}\ of Ag for Cu or Au. Likewise there is enormous scope for substitution on the [NEt$_4$]$^+$ site, and indeed variation in network topology \cite{Hill_2016,Schlueter_2005}. Whether or not it eventually proves possible to use XDW states to generate mesoscale chiral textures, our study has firmly established the ability of hybrid frameworks to support unconventional chiral order.

\begin{acknowledgments}
The authors gratefully acknowledge financial support from the E.R.C. (Grant 279705), and the E.P.S.R.C. (U.K.). We are extremely grateful for the award of beamtime to the Block Allocation Group award (MT9981) used to collect the Single Crystal Synchrotron X-ray diffraction data on the I19 beamline at the Diamond Light Source.
\end{acknowledgments}

\end{document}